# Large enhancement of transport critical current density of ex-situ PIT Ag/(Ba,K)Fe$_2$As$_2$ tapes achieved by applying cycles of cold deformation and heat treatment

Kazumasa Togano, Zhaoshun Gao, Akiyoshi Matsumoto, and Hiroaki Kumakura

National Institute for Materials Science, Tsukuba, Ibaraki 305-0047, Japan

## Abstract

We report that high transport critical current density $J_c$ exceeding $10^4$ A/cm$^2$ at 4.2 K and 10 T can be obtained with good reproducibility for Ag-sheathed (Ba,K)Fe$_2$As$_2$ (Ba-122) tapes by modifying the ex-situ powder-in-tube (PIT) method. The process involves cycles of the cold deformation and subsequent heat treatment. The intermediate cycles of flat rolling and heat treatment produce uniform fine-grained structure, resulting in an enhancement of transport $J_c$ in applied magnetic fields. However, the rolled tapes contain cracks running traverse to the tape axis, which still limit the current pass along the tape axis. We found that the application of uniaxial pressing with small reduction rate at the final stage is very effective to heal those cracks, resulting in an extraordinary increase of the transport $J_c$. We achieved the highest $J_c$ (at 4.2 K and 10 T) of $2.1\times10^4$ A/cm$^2$ among the PIT processed Fe-based wires and tapes reported so far. The $J_c$-$H$ curves measured at higher temperatures maintains a small filed dependence up to around 20 K, suggesting that the tapes are promising for applications at higher temperatures as well as in liquid helium.

The discovery of superconductivity at 26 K in LaFeAsO1-xFx[1] in early 2008 aroused enormous research on the iron-based superconductors and, to date, many families such as REOFeAs(1111-type)[2], LiFeAs(111-type)[3], BaFe$_2$As$_2$(122-type)[4], FeSe(11-type)[5] and the pnictides with perovskite-type blocking layer[6] have been found to show high temperature superconductivity up to 56 K by appropriate doping. In these, the K-doped BaFe$_2$As$_2$ and SrFe$_2$As$_2$ superconductors are most potentially useful for magnet applications due to their high $T_c$ value of ~39 K, $H_{c2}$ of over 50 T[7] and relatively small anisotropy.[8] Furthermore, it was reported that the transport $J_c$-$H$ curve maintains a small magnetic field dependence at temperatures up to 20 K[9], indicating that 122 superconducting wires are promising for magnet applications at the medium temperatures of cryogenic cooling or liquid hydrogen as well as in liquid helium.

In order to evaluate the potentiality for wire applications, the development of wire processing technique is essential. For Ba-122 and Sr-122, powder-in-tube (PIT) and coated conductor processes have been developed for wire or tape fabrication. Very high transport critical current density $J_c$ of well over $10^5$ A/cm$^2$ at 4 K and 10 T was reported for the coated conductor tape with Co-doped Ba-122 film, which was grown on a textured MgO buffer layer on metal substrate[10,11]. On the other hand, the transport $J_c$ of the PIT processed wires reported at an early stage of the development[12-17] was disappointingly low due to the weak link grain boundaries problem.[18] This might be arisen from the presence of numerous cracks and voids along the grain boundaries and/or the misorientation between the grains. Recently, there are several advances in the improvement of grain connectivity reported for the PIT processed 122 wires. Weiss et al[19] reported the transport $J_c$ of as high as ~8x10$^3$ A/cm$^2$ at 4.2 K and 10 T for the ex-situ PIT processed Cu/Ag/Ba-122 round wires, which was achieved by using a high quality precursor prepared by high energy ball milling and applying cold isotropic press (CIP) and hot isotropic press (HIP). The results indicate the importance of the quality of precursor and the densification of 122 core. On the other hand, Wang et al[20] and Gao et al[21] demonstrated that strong texture of c-axis alignment can be achieved by applying rolling process for Fe-sheathed Sr-122 wires and is effective to improve the grain connectivity as well as the densification. They reported the $J_c$ value (at 4.2 K and 10 T) of 1.7 x 10$^4$ A/cm,[22] which was the highest $J_c$ among the PIT processed iron based superconducting wires reported so far. Those results are encouraging the advances in the development of PIT process for Fe-based superconductors.

In addition to high transport $J_c$, the process should be simple and easy to be scaled up to the long length wire production, requiring little sophisticated equipments and processes. Like Bi-based high temperature superconductors, Ag looks the most suitable sheath material for 122 phase such as (Ba,K)Fe$_2$As$_2$ and (Sr,K)Fe$_2$As$_2$. This is because Ag does not react with 122 superconductors and is expected to serve as a stabilizing material of the conductor due to its low electrical resistivity. We have been studying on the PIT process for the fabrication of Ba-122 wires and tapes using a Ag single sheath.  In the previous paper[23], we demonstrated that the combined process of flat rolling and subsequent heat treatment is effective in enhancing the transport $J_c$.  The process is similar to that developed for the commercial production of Bi-2223 superconducting tapes in the early 1990s[24]. In the course of the development of PIT Bi-2223 tapes, several researches reported that the cold pressing is more effective for the enhancement of transport $J_c$ rather than flat rolling.[25,26]  The advantage of the pressing for current flow along the tape was directly evidenced by using magneto optical imaging.[27] It is natural that the pressing was also applied to the PIT process of $MgB_2$[28] and Ba-122[29] superconductors, which were discovered after the cuprate superconductors.  However, the $J_c$ improvement for Ba-122 was not so significant as expected from the result of Bi-2223 tapes.  In this paper, we demonstrate for the ex-situ PIT processed Ba-122 tapes that the uniaxial pressing is very effective in enhancing the transport $J_c$ when it is properly combined with flat rolling.

The precursors of $(Ba_{0.6}K_{0.4})Fe_2As_2$ were prepared from the Ba filings, K plates, Fe powder and As pieces. These element materials were mixed with the nominal composition of $(Ba_{0.6}K_{0.4})Fe_2As_{2.02}$ in an Ar atmosphere using a ball milling machine and put into a Nb tube of 6 mm outer diameter and 5 mm inner diameter for the heat treatment. The Nb tube was put into a stainless steel tube, whose both ends were pressed and sealed by arc welding in an Ar atmosphere.  The heat treatment was carried out at 900°C for 10 h followed by furnace cooling in a box furnace.  The $(Ba_{0.6}K_{0.4})Fe_2As_2$ precursor obtained has a fairly good quality as shown in the powder x-ray diffraction pattern and magnetization vs. temperature curve of **Fig. 1 (a) and (b)**, respectively.  The x-ray diffraction pattern was taken after pressing the powder on a glass plate.  All peaks are indentified as those from the Ba-122 structure and no peak from impurity phases was observed in the pattern.  The (00$l$) peaks from the basal plane show very high relative intensity compared to the random structure, indicating that each grain has a plate like morphology of single crystal and was forced to align parallel to the glass plate. The magnetization versus temperature curve measured by a SQUID magnetometer (MPMS) shows a sharp drop with the onset at around 37 K, which is another indication of the high quality of the precursor.

The precursor was then ground into powder using an agate mortar in a glove box filled with a high purity Ar gas.  The powder was packed into a Ag tube (outside diameter: 6 mm, inside diameter: 4 mm), which was subsequently groove rolled into a wire with the rectangular cross section of ~2 mm x ~2mm and then heat treated at 850˚C for 2 h.  After the first heat treatment, the wires were deformed into a tape form using a flat rolling machine initially into 0.6~0.7 mm in thickness followed by the second heat treatment at 850 ºC for 2 h and then into 0.4 ~0.5 mm in thickness followed by the third heat treatment at 850 ºC for 2 h . The tape was then cut into 35 mm in length and uniaxially pressed between two hardened steel dies.  By this uniaxial pressing, the thickness was reduced by 5~10% .  The pressed tape was then subject to the final heat treatment of 850 ºC for 10 h for sintering followed by the furnace cooling in a box furnace.  All heat treatment was carried out after putting the samples into a stainless steel tube, whose both ends were pressed and sealed by arc welding in an Ar atmosphere.

The critical current, $I_c$, measurement was carried out in liquid helium (4.2 K) using a 12 T superconducting magnet. The magnetic field was applied perpendicularly to the tape length and parallel to the tape surface. We found that the pressed samples have a very high $I_c$, even in strong magnetic fields above 10 T. **Figure 2 (a)** shows an example of voltage vs. applied current curves measured for the pressed tape of 0.4 mm in thickness. The transverse cross section of the tape is also shown in **Fig. 2 (b)**. The measurement at lower fields than 10 T was not carried out for this sample, because the current flow over 130 A causes thermal run away at the transition, resulting in sample damage. $I_c$ was determined using the voltage criterion of 1 µV/cm. Transport critical current density, $J_c$, was estimated by dividing the $I_c$ by the cross sectional area of the superconducting core, which was measured by using the image analysis of a laser optical microscope.

**Figure 3** shows the plots of the transport $J_c$ of three tapes with different final thickness fully processed by the cycle of cold deformation and heat treatment. For comparison, the $J_c$-$H$ curves of the wire with a rectangular cross section (~2 mm x ~2 mm) and the tape (0.4 mm thickness) subject to the final heat treatment of 850 ºC for 10 h without uniaxial pressing. The figure clearly indicates that the $J_c$ increases with the progression of the process. The $J_c$ of the wire is as low as ~$10^3$ A/cm$^2$ in applied magnetic fields. This is consistent to other papers[9,30,31], in which similar $J_c$ values were reported for Ba-122 and Sr-122 wires processed by the conventional PIT route. However, the $J_c$ is increased by applying the cycles of flat rolling and subsequent heat treatment, as we reported in the previous paper.[23] The enhancement is significant in the strong magnetic fields, while the $J_c$ at self field shows almost no change. As discussed later, the enhancement can be related to the smaller grain size of the tape. The most remarkable change in this figure is further large enhancement by the application of uniaxial pressing. All three pressed tapes show the $J_c$ over $10^4$ A/cm$^2$ at 10 T, indicating that the high $J_c$ is obtained with good reproducibility. The $J_c$ of 2.1 x $10^4$ A/cm$^2$ at 10 T observed for the 0.4 mm thickness tape is the highest among the PIT processed Fe based superconducting wires reported so far. It is expected from the extrapolation of $J_c$-$H$ curves that the $J_c$ of these three pressed tapes exceeds $10^5$ A/cm$^2$ in self field. We also carried out the $I_c$ measurement at high temperatures above 4.2 K using the temperature variable cryostat in a 15 T superconducting magnet. The result is shown in **Fig. 4**. It is noticed that the tape maintains small field dependence up to around 20 K, keeping a high $J_c$ value over $10^3$ A/cm$^2$ at 10 T. The curve shows a crossover with that of the PIT processed MgB$_2$ wire[32] at around 6 T. The result indicates that the tapes are promising for magnet applications at the medium temperatures of cryogenic cooling or liquid hydrogen as well as in liquid helium

In order to study the mechanism of $J_c$ enhancement, we investigated the microstructure change during the process. **Figure 5 (a) and (b)** show the optical micrographs observed on the cross sections of the wire and the tape, respectively. The wire was prepared by the groove rolling and subsequent heat treatment as mentioned above and the tape was fully processed by the cycles of cold deformation and heat treatment including uniaxial pressing at the final stage. Both were taken after the heat treatment. The microstructure of the wire shows nonuniformity in the grain size distributed widely from a few μm to more than 10 μm. On the other hand, the tapes subject to the cycles of cold deformation and heat treatment show a very uniform structure on the whole of the cross section composed of grains whose size is a few μm, as shown in Fgi. 5 (b). We consider that the cycles of flat rolling and subsequent heat treatment serves for grain refining, breaking up a large grain into smaller grains and resulting in the observed uniform fine-grained structure. Finer grain structure is expected to bring about stronger pinning force by the grain boundary and higher $J_c$ in applied magnetic fields. It is interesting that the grains have almost equiaxed morphology, although the starting powder is composed of single crystals. This is consistent with the x-ray diffraction pattern taken on the core surface shown in **Fig. 6 (a)**, in which the relative intensity of (00$l$) peaks is not so high as observed in Fig. 1 (a). This is different from the PIT processed Bi-2223 tape, in which the cycles of the flat rolling and heat treatment produce strong c-axis alignment due to the larger anisotropic morphology of Bi-2223 crystal[24]. The result is also contrast to the Fe sheathed PIT Sr-122 tape[20-22], in which strong texture was observed similarly to the Bi-2223 tape. It is considered that the difference is caused by the different sheath material and different processing parameters such as reduction rate. In **Fig. 6 (b)**, we showed the magnetization vs. temperature curves measured for the core of the pressed tape, which shows a sharp transition with the onset at around 35 K.

The structure change by the uniaxial pressing is more interesting to be investigated, because the increment of $J_c$ is much larger than the flat rolling. However, despite careful microstructure observations before and after the pressing, we could not see any clear difference for the matrix structure at microscopic levels, both tapes being composed of fine Ba-122 grains as shown in Fig. 5(b). However, we could see an apparent difference in crack structures. The observation was performed on the cross section parallel to the tape plane after the polishing. Cracks running from the interface with the Ag sheath to inside the core were frequently observed for the as rolled tapes as shown in **Fig. 7 (a)**, while they are seldom observed for the as pressed tapes. We also observed smaller cracks inside the core in both tapes as shown in **Fig. 7 (b) and (c)**. They run traverse to the tape length in the as rolled tape, while they run parallel to the tape length in the as pressed tape. We conclude that the observed difference in crack direction is responsible for the large difference in transport $J_c$, because the cracks run traverse to the tape length reducing the effective cross sectional area for the current flow along the tape length. This is quite similar to the case of Bi-2223 tape.[25-27] The optimum reduction rate for each cold deformation would be determined by the balance between the improvement of density and the initiation of microcracks which cannot be healed by subsequent heat treatment. We are now performing more systematic investigations to optimize the various processing parameters such as the number of cycle, reduction rate, heat treatment conditions and so on.

In summary, high transport Jc exceeding $10^4$ A/cm$^2$ at 4.2 K and 10 T was obtained with good reproducibility for Ag-sheathed ex-situ PIT Ba-122 tapes by applying the cycles of the cold deformation and subsequent heat treatment. The intermediate cycles of flat rolling and annealing produce uniform fine-grained structure, which might be beneficial to strengthen the flux pinning by the grain boundary. We also found that the addition of uniaxial pressing with small reduction rate at the final stage is very effective to heal the cracks formed by the previous rolling deformation, which run traverse to the tape axis. The transport $J_c$ is further increased by one order of magnitude by applying this uniaxial pressing. We achieved the highest transport $J_c$ (at 4.2 K and 10 T) of $2.1\times10^4$ A/cm$^2$ among the PIT processed Fe-based superconducting wires reported so far. We also found that the $J_c$-$H$ curves maintains a small filed dependence up to the temperatures of around 20 K, suggesting that the tapes are promising for magnet applications at the medium temperatures of cryogenic cooling or liquid hydrogen as well as in liquid helium. The process is simple using single sheath of Ag and the results obtained in this work verify that the PIT method is promising for the practical production of Fe-based superconducting wires with high transport $J_c$ like Bi-based cuprate and MgB$_2$ superconducting wires.

**Acknowlegdements**

This work was supported by the Japan Society for the Promotion of Science (JSPS) through its "Funding Program for World-Leading Innovative R&D on Science and Technology (FIRST) Program. We acknowledge Dr. H. Fujii and Mr. S. J. Ye of the National Institute for Materials Science for their assistance in $I_c$ measurement.

**References**

[1]Y. Kamihara, T. Watanabe, M. Hirano, and H. Hosono: J. Am. Chem. Soc. **130**, 3296(2008).

[2]X. H. Chen, T. Wu, G. Wu, R. H. Liu, H. Chen, and D. F. Fang: Nature **453**, 761(2008).

[3]X. C. Wang, Q. Q. Liu, Y. X. Lu, W. B. Gao, L. X. Yang, R. C. Yu, F. Y. Li, and C. Q. Jin: Phys. Rev. B **78**, 060505-1-4 (2008).

[4] M. Rotter, M. Tegel, and D. Johrendt: Phys. Rev. Lett. **101**, 107006 (2008).

[5]F. C. Hsu, J. Y. Luo, K. W. The, T. K. Chen, T. W. Huang, P. M. Wu, Y. C. Lee, Y. L. Huang, Y. Y. Chu, D. C. Yan and M. K. Wu: Proc. Natl. Acad. Sci. U.S.A. **105**, 14262 (2008).

[6]H. Ogino, Y. Matsumura, Y. Katsura, K. Ushiyama, S. Horii, K. Kishio, and J. Shimoyama: Supercond. Sci. Technol. **22**, 75008 (2009).

[7]M. Putti, I. Pallecchi, E. Bellingeri, M. R. Cimberle, M. Tropeano, C. Ferdeghini, A. Palenzona, C. Tarantini, A. Yamamoto, J. Jiang, J. Jaroszynski, F. Kametani, D. Abraimov, A. Polyanskii, J. D. Weiss, E. E. Hellstrom, A. Guverich, D. C. Larbalestier, R. Jin, B. C. Sales, A. S. Sefat, M. A. McGuire, D. Mandrus, P. Cheng, Y. Jia, H. H. Wen, S. Lee, and C.B. Eom: Supercond. Sci. Technol. **23**, 034003 (2010).

[8]M.A.Tanatar, N. Ni, C. Martin, R.T. Gordon, H. Kim, V.G. Kogan, G.D. Samolyuk, S.L. Bud'ko, P.C. Canfield and R. Prozorov, Phys. Rev. B **79**, 094507.

[9]A. Matsumoto, K. Togano and H. Kumakura, Supercond. Sci. Technol **25**, 125010(2012).

[10]T. Katase, H. Hiramatsu, V. Matias, C. Sheehan, Y. Ishimaru, T. Kamiya, K. Tanabe and H. Hosono, Appl. Phys. Lett. **98**, 242510(2011).

[11]S. Trommler, J. Haenisch, V. Matias, R. Huehne, E. Reich, K. Iida, S. Haindl, L. Schultz and B. Holzapfel, Supercond. Sci. Technol. **25**, 084019.

[12]Y. Mizuguchi, K. Deguchi, S. Tsuda, T. Yamaguchi, H. Takeya, H. Kumakura, and Y. Takano : Appl. Phys. Express **2** (2009) 083004.

[13]L. Wang, Y. P. Qi, D. L. Wang, X. P. Zhang, Z. S. Gao, Z. Y. Zhang, Y. W. Ma, A. Awaji, G. Nishijima, and K. Watanabe: Physica C **470** (2010) 183.

[14]Y. P. Qi, L. Wang, D. L. Wang, Z. Y. Zhang, Z. S. Gao, X. P. Zhang, and Y. W. Ma: Supercond. Sci. Technol. **23** (2010) 055009.

[15]L. Wang, Y. P. Qi, D. L. Wang, Z. S. Gao, X. P. Zhang, Z. Y. Zhang, C. L. Wang, and Y. W. Ma: Supercond. Sci. Technol. **23** (2010) 075005.

[16]M. Fujioka, T. Kota, M. Matoba, T. Ozaki, Y. Takano, H. Kumakura, and Y. Kamihara, Appl. Phys. Exoress **4** (2011) 063102.

[17]K. Togano, A. Matsumoto, and H. Kumakura, Appl. Phys. Express **4** (2011) 043101.

[18]J.H. Durrell, C-B Eom, A. Guverich, E.E. Hellstrom, C. Tarantini, A. Yamamoto and D.C. Larbalestier, Rep. Prog. Phys. 74, 124511(2011).

[19]J.D. Weiss, C. Tarantini, J. Jiang, F. Kametani, A.A. Polyanskii, D.C. Larbalestier and E.E. Hellstrom, Nature Materials **11**, 682(2012).

[20]L. Wang L, Y.P. Qi, X.P. Zhang, D.L. Wang, Z.S. Gao, C.L. Wang, C. Yao and Y.W. Ma, Physica C **471**, 1689(2011).

[21]Z.S. Gao, L. Wang L, Y. Chao Y, Y.P. Qi, C.L. Wang, X.P. Zhang, D.L. Wang, C.D. Wang and Y.W. Ma, Appl. Phys. Lett. **99**, 242506(2011).

[22]Z.S. Gao, Y.W. Ma, C. Yao, X.P. Zhang, C.L. Wang, D.L. Wang, S. Awaji and K. Watanabe, Scientific Reports, **2**, 998(2012).

[23]K. Togano, Z.S. Gao, A. Matsumoto and H. Kumakura, accepted for the publication in Supercond. Sci. Technol.

[24]R. Flukiger, T. Graf, M. Decroux, C. Groth and Y. Yamada, IEEE Trans. Mag. **2**, 1258(1991).

[25]Q. Li, K. Brodersen, H.A. Hjuler and T. Freltoft, Physica C **217**, 360(1993)

[26]G. Grasso, A. Jeremie and R. Flukiger, Supercond. Sci. Technol. **8**, 827(1995).

[27]J.A. Parrell, A.A. Polyanskii, A.E. Pashitski and D.C. Larbalestier, Supercond. Sci. Technol. **9**, 393(1996).

[28]R. Flukiger, M.S.A. Hossain and C. Senatore, Supercond. Sci. Technol. **22**, 085002(2009).

[29]Y. Ding, G.Z. li, Y. Yang, C.J. Kovacs, M.A. Susner, M.D. Sumption, Y. Sun, J.C. Zhuang, Z.X. Shi, M. Majoros and E.W. Collings, Physica C **483**, 13(2012).

[30]Y.W. Ma, L. Wang, Y.P. Qi, Z.S. Gao, D.L. Wang and X.P. Zhang, IEEE Trans. Appl. Supercond. **21**, 2878(2011).

[31]Q.P. Ding, T. Prombood, Y. Tsuchiya, Y. Nakajima and T. Tamegai, Supercond. Sci. Technol. **25**, 035019(2012).

[32] H. Kumakura, A. Matsumoto, T. Nakane and H. Kitaguchi, Physica C **456,** 196(2007)

**Figure Captions**

Figure 1   (a) Poweder x-ray diffraction pattern and (b) magnetization vs. temperature curves for the $(Ba_{0.6}K_{0.4})Fe_2As_2$ precursor.

Figure 2 (a) Voltage vs. applied current curves measured for the tape (0.4 mm thickness) processed by the cycles of cold deformation and subsequent heat treatment with uniaxial pressing.   (b) The cross section of the tape.

Figure 3   Plots of transport critical current density, $J_c$, as a function of magnetic field, $H$, measured for three samples of the tape with different thickness fully processed by the cold deformation and subsequent heat treatment with uniaxial pressing at the final stage.   For comparison, the $J_c$-$H$ curve of the wire processed without rolling and pressing and that of the tape processed without pressing are shown in the figure.

Figure 4   Plots of transport critical current density, $J_c$, as a function of magnetic field, $H$, measured at high temperatures above 4.2 K.   The sample is fully processed by the cycles of cold deformation and subsequent heat treatment with uniaxial pressing at the final stage.   The $J_c$-$H$ curve at 20 K of PIT processed $MgB_2$ wire is shown for comparison[32]

Figure 5   Optical micrographs observed on the cross sections of the wire (a) and tape (b).   The wire was prepared by the groove rolling and subsequent heat treatment.   (b) The tape was fully processed by the cycles of cold deformation with uniaxial pressing at the final stage.

Figure 6 (a) X-ray diffraction (XRD) pattern and (b) magnetization vs. temperature curves for the tape fully processed by the cycles of cold deformation and subsequent heat treatment.   The XRD measurement was carried out on the surface of the core pealed off from the Ag sheath..

Figure 7   The cracks observed on the core of the as rolled ((a) and (b)) and as pressed (c) tapes. The observation was performed on the cross section parallel to the tape plane after the polishing.

Fig. 1

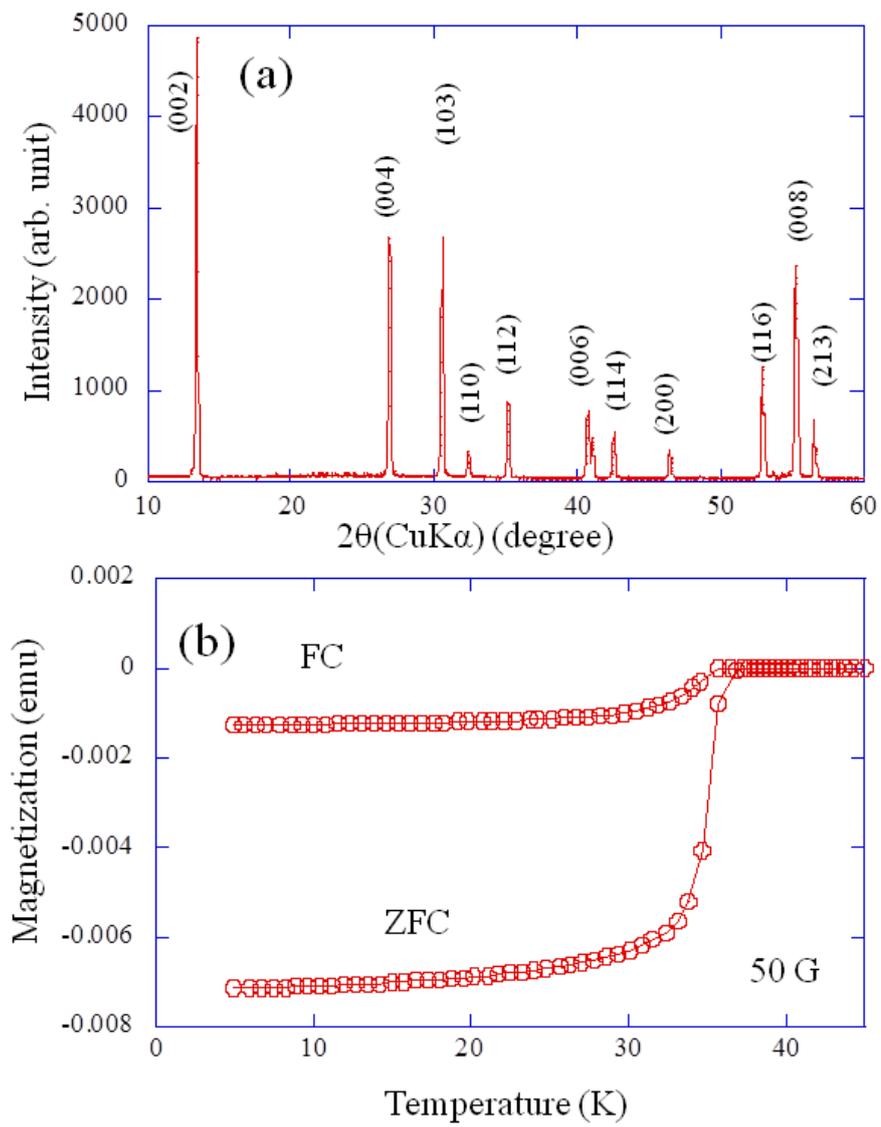

Fig. 2

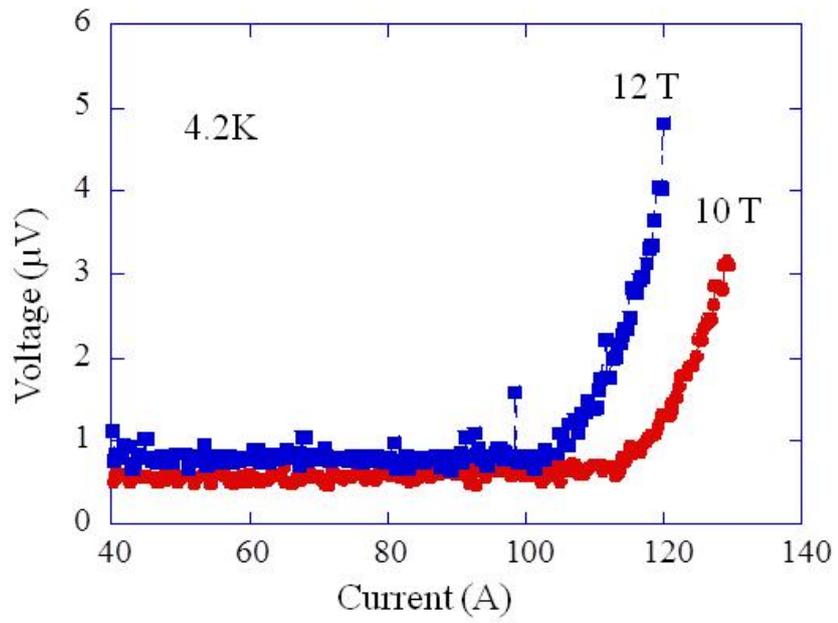

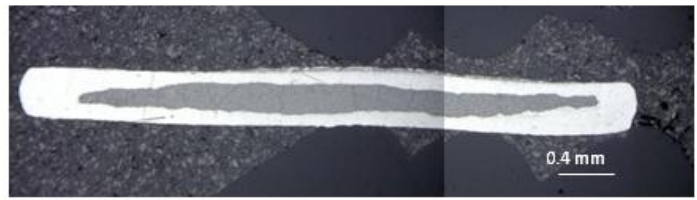

Fig. 3

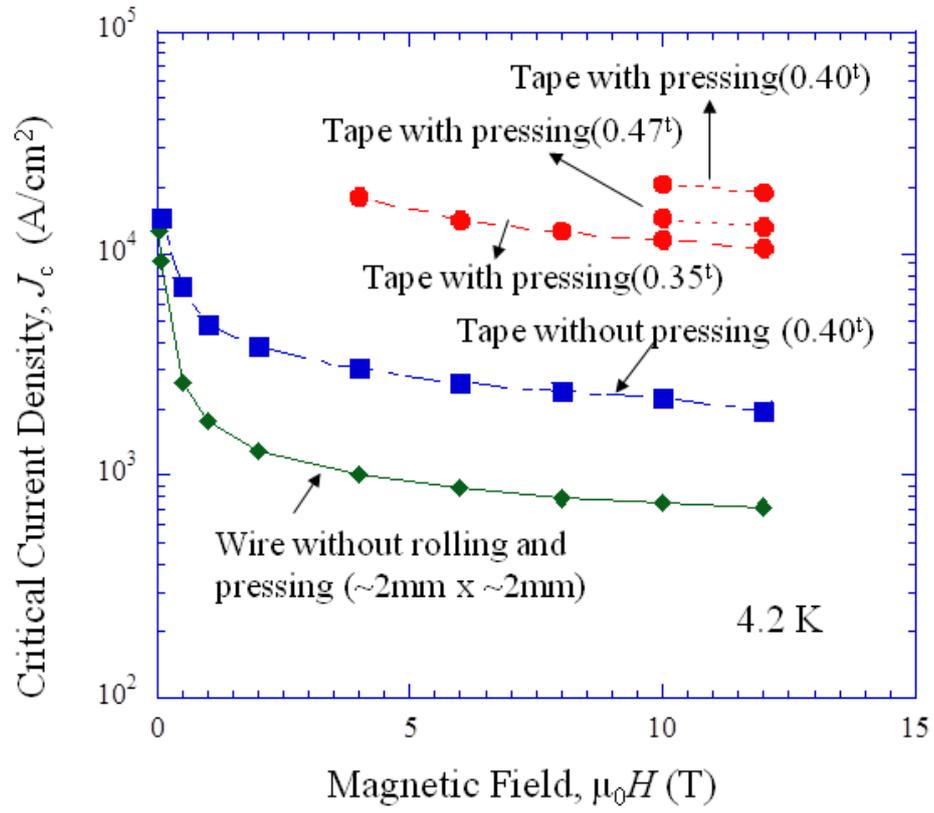

Fig. 4

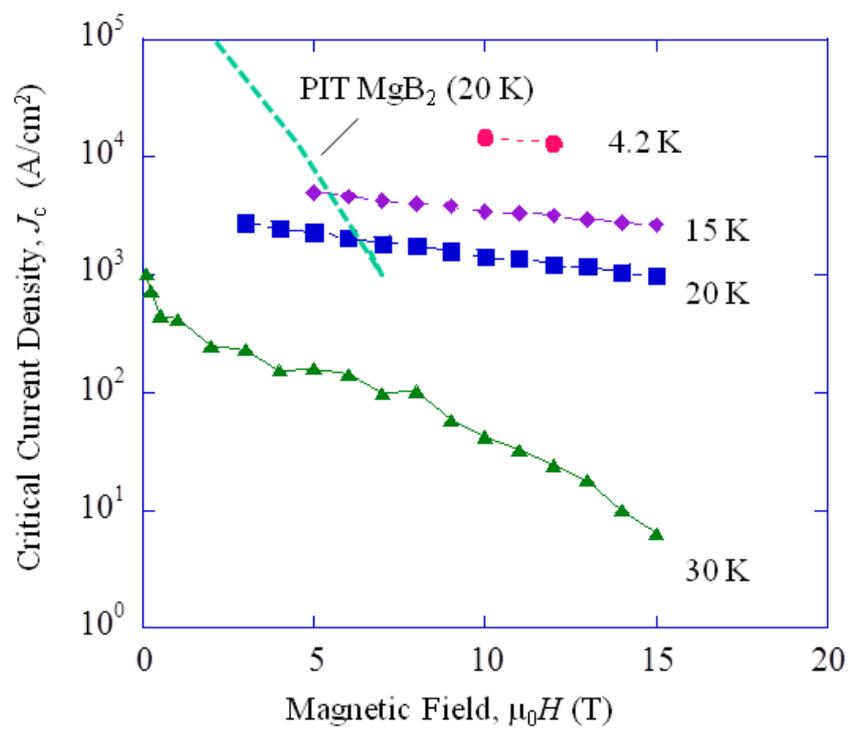

Fig. 5

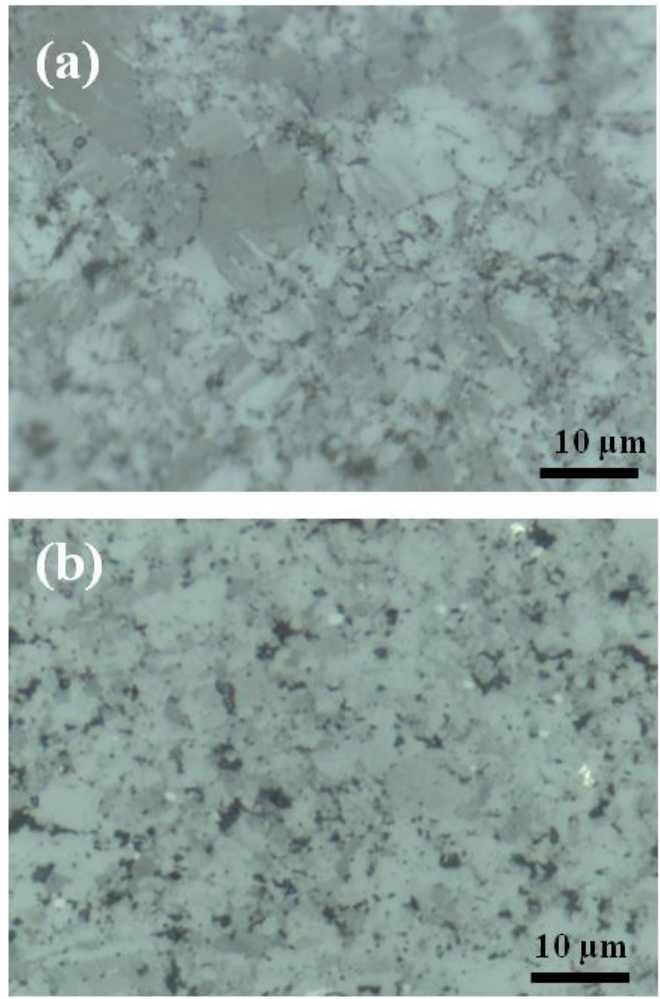

Fig. 6

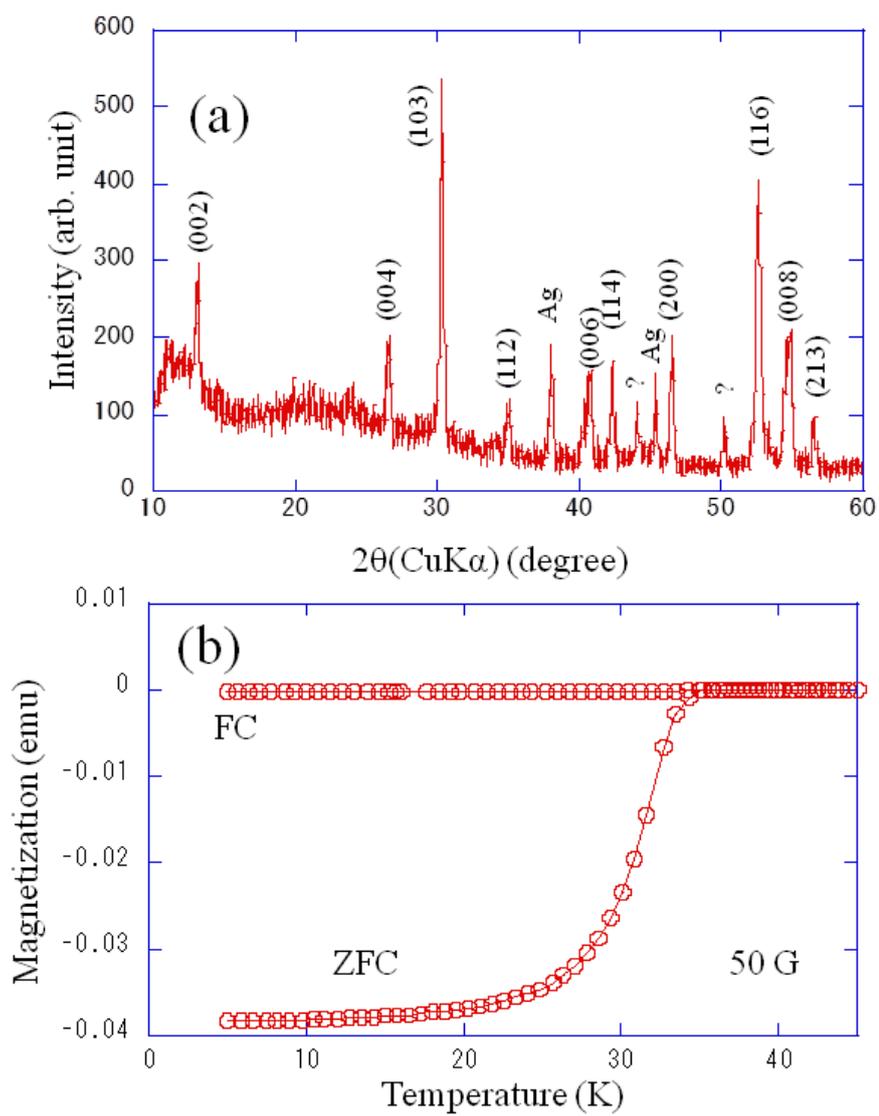

Fig. 7

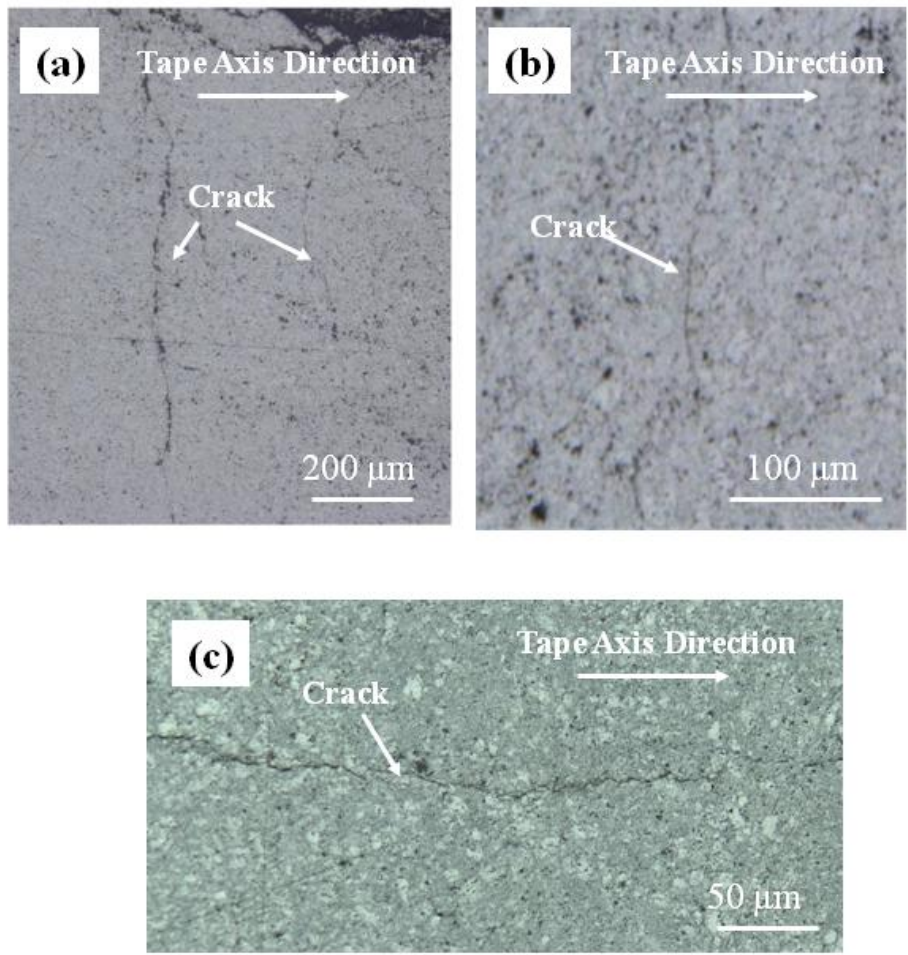